\documentstyle[referee]{mn}

\newcommand{\reference}{\bibitem}

\def\araa{ARAA}
\def\aap{A\&A}
\def\apj{ApJ}
\def\plotone#1{\centering \leavevmode
\epsfxsize=\columnwidth
\epsfbox{#1}}

\def\beq{\begin{equation}}
\def\eeq{\end{equation}}
\def\bey{\begin{eqnarray}}
\def\eey{\end{eqnarray}}

\def\kpc{\,{\rm {kpc}}}

\def\chisq{\chi^2}

\def\tE{t_{\rm E}}
\def\fs{f_{\rm S}}

\def\rEt{\tilde{r}_{\rm E}}

\def\rE{r_{\rm E}}

\def\u0{u_0}

\def\Ds{D_{\rm S}}
\def\Dl{D_{\rm L}}
\def\mI0{m_{I,0}}

\def\pirel{\pi_{\rm rel}}
\def\sc5{sc5\_2859}

\input epsf
\begin{document}

\title[The OGLE-II event \sc5 -- An example of disk-disk microlensing]
{The OGLE-II event \sc5 -- An example of disk-disk microlensing}

\author[M. C. Smith]
{Martin C. Smith$^1$
\thanks{e-mail: msmith@jb.man.ac.uk}
\\
\smallskip
$^{1}$Univ. of Manchester, Jodrell Bank Observatory, Macclesfield,
Cheshire SK11 9DL, UK
}
\date{Accepted ........
      Received .......;
      in original form ......}

\pubyear{2002}

\maketitle
\begin{abstract}
We present a new long-duration parallax event from the OGLE-II
database, \sc5, which has the second longest time-scale ever
identified ($\tE = 547.6^{+22.6}_{-7.8}$ days). We argue that both the
lens and source reside in the Galactic disk, making event \sc5 one of
the first confirmed examples of so-called disk-disk microlensing. We find that
the source star is most probably located at a distance
of $D_{\rm S}\sim 2$ kpc, and from this we conclude that the lens is
unlikely to be a main-sequence star due to the strict limits that can be
placed on the lens brightness.
A simple likelihood analysis is carried out on the lens mass,
which indicates that the lens could be another candidate stellar
mass black hole.
We recommend that spectroscopic observations of the source be
carried out in order to constrain the source distance, since this is
the main source of uncertainty in our analysis.
In addition, we briefly discuss whether there appears to be an excess
of long duration microlensing events in the OGLE-II catalogue.
\end{abstract}


\section{Introduction}

It is nearly ten years since the first gravitational microlensing
event was detected toward the Galactic bulge (Udalski et al. 1993).
Since then, microlensing has proved to be a useful tool for many
astrophysical applications
(for a review of Galactic microlensing, see Paczy\'nski 1996). In the
past, however, studies of microlensing statistics have been limited by
the small number of detected events. This situation is improving as
new microlensing projects begin operation, e.g., the 
OGLE-III\footnote{http://www.astrouw.edu.pl/\~{}ogle/ogle3/ews/ews.html}
project, which is hoping to detect up to 1000 per year.

Two ways that microlensing statistics can be used to investigate
Galactic models are through the study of the observed microlensing
optical depth and the distribution of event time-scales. The
optical depth toward the Galactic bulge has been studied by many
collaborations (for example, Alcock et al. 2000 and Sumi et al. 2002)
and the resulting values are significantly larger than predicted
estimates (see, for example, Binney, Bissantz \& Gerhard 2000; Evans
\& Belokurov 2002; Klypin, Zhao \& Somerville 2002), although recent
work has questioned the significance of this discrepancy (Popowski
2002; Afonso et al. 2003).

The distribution of event time-scales, which can be used to
investigate the mass spectrum of lensing objects (e.g. Han \& Gould
1996; Peale 1998), also seems to
disagree with predicted estimates; it has long been suspected that
microlensing searches appear to be identifying an unexpectedly large
proportion of long-duration events. An important early study into the
distribution of microlensing time-scales was carried out by Han \&
Gould (1996; see also Zhao, Rich \& Spergel 1996).
They found that in one-year of microlensing observations from the OGLE
and MACHO collaborations 8\% of events had time-scale greater than 70
days, whereas the largest fraction predicted by their theoretical
models was 2\%.
Bennett et al. (2002a) analysed a more recent distribution of events
timescales from MACHO data and found a similar excess of long
duration events. Although no systematic analysis of event time-scales
has been performed for the full 4-year OGLE-II catalogue, a
preliminary investigation using the first 3-years of data also
indicates the existence of an excess (Udalski et al. 2000).

To date, two microlensing events have been identified with
time-scales greater than one year (OGLE-1999-BUL-32/MACHO-99-BLG-22,
Mao et al. 2002, Bennett et al. 2002b; OGLE-1999-BUL-19, Smith et
al. 2002). We will introduce a third event, \sc5, in this paper. It
has been proposed that some of these extreme 
long-duration events could be due to lensing by massive stellar
remnants, e.g. black holes (Agol et al. 2002; Bennett et al. 2002a),
although it is unlikely that this hypothesis could account for every
long-duration event, for example OGLE-1999-BUL-19 (Smith et al. 2002).

In this paper we present the analysis of event \sc5. We begin by
describing the observational data (Section \ref{data}), before
proceeding to fit the event with both the standard and parallax
microlensing models (Section \ref{model}). We then investigate whether
any constraints can be placed on the lens mass; we first attempt to
utilise finite source size effects (Section \ref{fss}), before
considering more general arguments based on the relative transverse
velocity of the lens and limits on the lens brightness (Section
\ref{lhood}). In Section \ref{fss} we also discuss the possible
location and spectral type of the source star.
Section \ref{disc} contains a brief discussion regarding the nature of
event \sc5 and the possibility that there are an excess of long
duration microlensing events in the OGLE-II catalogue. This section
also mentions possible scientific returns from such long duration
events.
We conclude with a summary (Section \ref{conc}).

\section{Observational data}
\label{data}

The event \sc5 was identified during the second phase of the OGLE
project (in which it was named BUL\_SC5 244353; Udalski et al. 2000) toward
the Galactic bulge. The observations were carried out with the 1.3 m Warsaw
telescope at the Las Campanas Observatory, Chile, which is operated by
the Carnegie Institution of Washington. The instrumentation of the
telescope and CCD camera are described in detail by Udalski, Kubiak, \&
Szyma{\'n}ski(1997). The position of the source is RA=17:50:36.09 and
Dec=$-$30:01:46.6 (J2000), corresponding to Galactic coordinates
$l=359.6235^{\circ}$ and $b=-1.4930^{\circ}$. This event is also
included in the catalogue of Wo\'zniak et al. (2001), which employs
the Difference Image Analysis method of data reduction (Alard \&
Lupton 1998). This method generally appears to result in greater
accuracy compared to the classical point spread function approach
(Schechter, Mateo \& Saha 1993), and so we have chosen to use the
Difference Image Analysis data in the following analysis\footnote{
This analysis uses the full 4-year Difference Image Analysis data,
generously provided for our use by the OGLE collaboration. The partial
3-year Difference Image Analysis data, along with the basic
calibration data for this event, is publicly available online from
http://astro.princeton.edu/\~{}wozniak/dia/lens.}.
Unfortunately, this event was only detected after the peak in
magnification had already occurred, which means that less than half of
the full light curve is available for analysis.

\begin{figure}
\plotone{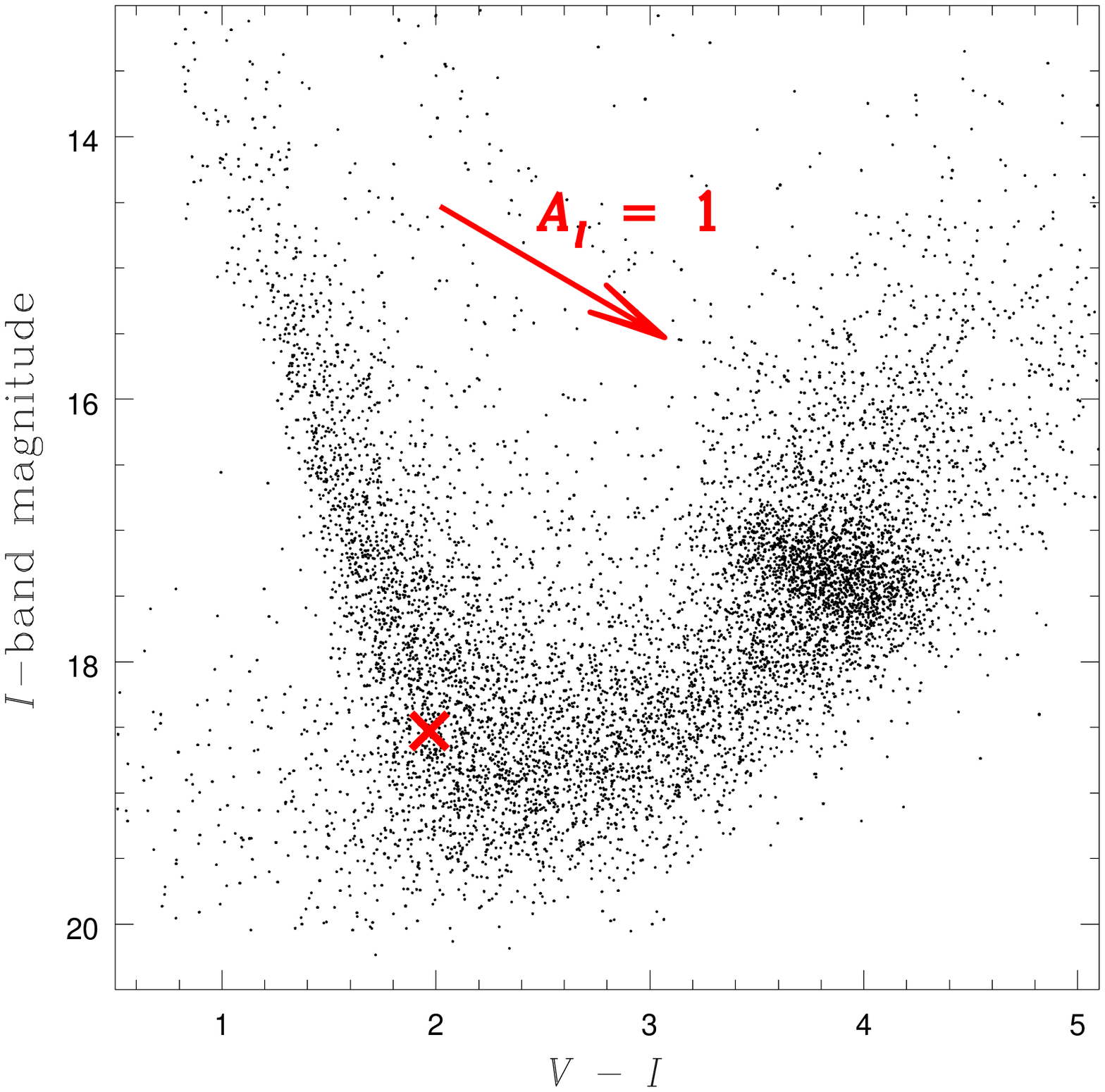} 
\caption{The colour-magnitude diagram for stars within 5 arcmin around
\sc5. The position of the lensed star is denoted by a cross and the
arrow denotes the reddening vector. From this figure
it appears that the source is located in the main sequence branch and
we believe that the source probably resides in the disk (see Section
\ref{fss}). 
The red clump giant region is clearly visible at $I\approx17.4$,
$(V-I)\approx3.8$, which highlights that this field is subject to a
large amount of interstellar extinction owing to the fact that the
field of view is very close to the Galactic plane ($b\approx-1.5^\circ$
for \sc5). This figure was produced using the $VI$ photometric maps of
the Galactic bulge presented in Udalski et al. (2002) and available
online from http://bulge.princeton.edu/\~{}ogle/ogle2/bulge\_maps.html.}
\label{fig:CMD}
\end{figure}

The baseline magnitude for the source is given by $I = 18.53 \pm 0.02$
mag\footnote{
It should be noted that there is a discrepancy in the baseline
magnitude of the source between the two photometric methods: the full
4-year Difference Image Analysis data have a baseline of $I = 18.53
\pm 0.02$ mag, but the 3-year classical point spread function data
have a baseline of $I = 18.86 \pm 0.02$ mag. However, the conclusions
regarding the nature of event \sc5 remain unchanged if the classical
point spread function data is used.}, and the colour is $(V-I)
= 1.97 \pm 0.02$ mag.
In Fig. \ref{fig:CMD} we present the colour-magnitude diagram for the
field around \sc5. From this figure it can be seen that the field
around \sc5 is subject to a large amount of interstellar extinction,
owing to the fact that the source is located very close to the
Galactic plane. It appears that the lensed star is located in the main
sequence branch of the colour-magnitude diagram. We will return to the
issue of the possible source location and spectral type in Section
\ref{fss}.


\section{Model fitting}
\label{model}

We initially fit the event with the standard microlensing model, which
assumes that the observer, lens and (point) source all move with
constant velocities. The magnification is given by (see, for example,
Paczy\'nski 1986),
\beq \label{amp}
A(t) = {u^2+2 \over u \sqrt{u^2+4}},~~
u(t) \equiv \sqrt{\u0^2 + \tau(t)^2},
\eeq
where $\u0$ is the impact parameter (in units of the Einstein radius)
and, 
\beq \label{tau}
\tau(t) = {t-t_0 \over \tE},
\eeq
with $t_0$ being the time of the closest approach (i.e., maximum,
magnification), and $\tE$ the event time-scale. The time-scale is
defined such that,
\beq
\tE = {\rE \over v} = {\tilde{r}_{\rm E} \over \tilde{v}},
\eeq
where $\rE$ is the lens' Einstein radius, $v$ is the lens' transverse
velocity relative to the observer-source line of sight, and
$\tilde{r}_{\rm E}$ and $\tilde{v}$ are the values of these two
quantities projected onto the observer plane. Therefore, $\tE$
corresponds to the time it takes for the lens to move a distance equal
to its Einstein radius\footnote{An alternate definition for the
time-scale is sometimes employed (particularly by the MACHO
collaboration), which uses $\hat{t}$, the Einstein {\it diameter}
crossing time, i.e. $\hat{t} = 2 \tE$.}.

The Einstein radius projected onto the observer plane is related to the
mass according to the following equation,
\beq \label{eq:rE}
\tilde{r}_{\rm E} = \sqrt{4 G M \Ds x \over {c^2 (1-x)}},
\eeq
where $M$ is the lens mass, $\Ds$ the distance to the source and
$x=\Dl/\Ds$ is the ratio of the distance to the lens and the distance
to the source. This shows the well-known degeneracy inherent in the
standard microlensing formalism; the quantities $\tilde{v}$, $M$ and
$x$ cannot be determined uniquely from a given microlensing light
curve, even if the source distance is known.

The Difference Image Analysis flux is given by,
\beq
F(t) = F_0 \left( \fs \left[ A(t) - 1 \right] + 1 \right) - F_{\rm Ref},
\eeq
where $F_0$ is the total baseline flux from the source plus any
blended star(s), if present, $\fs$ is the ratio of the baseline source
flux to the total baseline flux (i.e. a measure of the blending), and
$F_{\rm Ref} = 416.15$ is the flux of the reference image. All the fluxes here
are in units of 10\,ADU and can be converted into $I$-band magnitudes
using the following transformation (given in Wo\'zniak et al. 2001),
\beq
I(t) = I_{\rm Ref} - 2.5\,{\rm log}_{10}
\frac{F(t)}{F_{\rm Ref}},
\eeq
where $I_{\rm Ref}=16.76$ is the $I$-band magnitude of the reference
image. Note that the reference image is brighter than the baseline
magnitude for \sc5.

\begin{figure}
\plotone{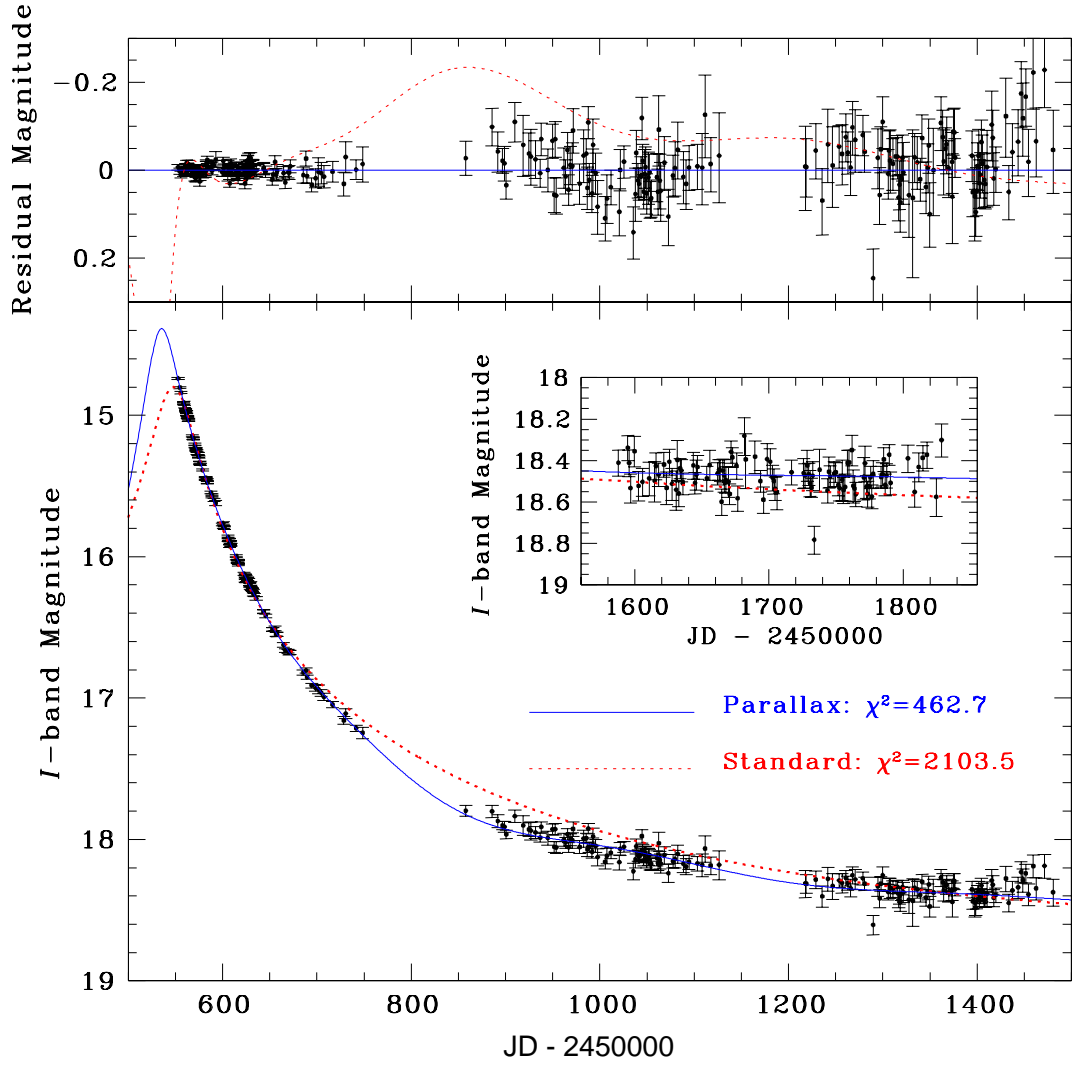} 
\caption{The {\it I}-band light curve for the OGLE-II event \sc5 from
Difference Image Analysis.
The best-fit standard and parallax 
models are given by the dotted and solid lines respectively. The upper
panel shows the residual magnitude (the observed data points with the
parallax model subtracted), and the inset shows the 4th
season. Clearly the standard model is unable to
reproduce the observed behaviour, but the parallax model provides a
good fit with a $\chi^2$ per degree of freedom of 1.23.}
\label{lc:sc5}
\end{figure}

We fit the light curve with the above five parameters for the standard
model, i.e., $t_0$, $\tE$, $u_0$, $F_0$, $\fs$. The best-fit parameters are
given in Table \ref{tb:sc5}, and the corresponding light curve is
shown in Fig. \ref{lc:sc5}. Clearly this model is unable to provide a
suitable fit for the data, with the best standard $\chisq$ per degree of
freedom greater than 5.

\begin{table*}
\begin{center}
\caption{The best standard model (first row) and the best parallax
model (second row) for \sc5. The final column shows the $\chisq$
and number of degrees of freedom (dof) for each model. The
parameters are explained in Section \ref{model}.}
\label{tb:sc5}
\scriptsize
{
\begin{tabular}{ccccccccc}
Model & $t_0$ & $\tE$ (day) & $\u0$ & $F_0$
& $\fs$ & $\psi$ (radians) & $\rEt$ (au) &
$\chisq$/dof \\
\hline
S &
$    547.80 \pm 0.20 $ &
$    815.4^{+ 7.5}_{-7.4} $ &
$     -0.02767 \pm 0.00028 $ &
$     70.17 \pm 0.54 $ &
$      1.00000^{+ 0}_{-0.00042} $ &
--- & --- &
2103.5 / 377
\\ \\
P & 
$    542.20 \pm 0.87 $ &
$    547.6^{+22.6}_{-7.8} $ &
$      0.0645^{+ 0.0046}_{-0.0045} $ &
$     81.83 \pm 0.54 $ &
$      1.000^{+ 0}_{-0.065} $ &
$      0.0477^{+ 0.0021}_{-0.0020} $ &
$     10.84^{+ 0.44}_{-0.41} $ &
462.7 / 375
\end{tabular}
}
\end{center}
\end{table*}

Since the duration of this event is longer than one year, the next
logical step is to fit the light curve with a model that incorporates
the parallax effect (Gould 1992). This effect, which arises when the
Earth's motion around the Sun is considered, is described in detail in
Soszy\'nski et al. (2001); see also Alcock et al. (1995), Dominik
(1998). It requires two additional parameters, the Einstein radius
projected onto the observer plane, $\rEt$, and an angle in the
ecliptic plane, $\psi$, describing the orientation of the lens
trajectory (given by the angle between the heliocentric ecliptic
$x$-axis and the normal to the trajectory).

The parallax model produces a drastic reduction in $\chisq$. The best-fit
parameters are given in Table \ref{tb:sc5}, and the corresponding
light curve is shown in Fig. \ref{lc:sc5}. The $\chisq$
per degree of freedom is 1.23, which indicates that the parallax model
provides a reasonable fit to the data.

We also fit this event with a slight variation on the above
parallax model. In the above model we describe the geometrical
properties in the ecliptic plane and then project these quantities
into the lens plane. However, since the ecliptic plane intersects the
Galactic bulge, this can lead to the projection being almost singular. 
Therefore, to avoid this potential singularity, one can instead take
the more conventional approach and describe the geometrical properties
in the plane perpendicular to the line-of-sight (see, for example,
Dominik 1998, Alcock et al. 1995); to describe the lens plane 
coordinate system we form a right-handed set with the $x$-axis chosen
to correspond to the North Ecliptic Pole projected onto the lens plane
and the $z$-axis chosen to be the observer-source line-of-sight, which
implies that the $y$-axis corresponds to the intersection of the lens
plane and the Ecliptic plane. We find that the best-fit parameters are
practically identical and therefore we do not present them
here. However, for this coordinate system, the angle describing the
relative lens-source trajectory is found to be
$\theta=-35.96^{+0.66}_{-1.20}$ degrees, where $\theta$ is the angle
between the trajectory and the $x$-axis (measured from the positive
$x$-axis towards the positive $y$-axis).

From Table \ref{tb:sc5}, the parallax parameters that provide
information regarding the lens properties are,
\beq
\rEt = 10.84^{+ 0.44}_{-0.41}~{\rm au},~~\tE =
547.6^{+22.6}_{-7.8}~{\rm days},
\eeq
which implies that the transverse velocity of the lens
relative to the source, projected onto the observer plane is,
\beq
\label{eq:v}
\tilde{v} = \frac{\rEt}{\tE} = 34.2^{+1.5}_{-1.9}~\rm{km s}^{-1}.
\eeq
If we convert the direction of $\tilde{v}$ from the lens plane
($\theta=-36.0$ degrees) into the plane perpendicular to the line-of-sight to
the Galactic centre, we find that $\tilde{v}$ is directed almost
parallel to the Galactic plane in the direction of rotation
($\theta_{\rm G}=-6.1^{+0.7}_{-1.2}$ degrees, where $\theta_{\rm G}$ is
measured from the Galactic plane towards the North Galactic Pole).
In this aspect \sc5 is similar to the strong parallax events
presented in Bennett et al. (2002a), all of which had $-90^\circ <
\theta_{\rm G} < 0 ^\circ$.

This timescale of 548 days is the second longest time-scale ever
identified (after the event OGLE-1999-BUL-32/MACHO-99-BLG-22; see Mao
et al. 2001, Bennett et al. 2002b). In addition, the value of $\rEt$
is also unusually large, which implies that this event could be
another black-hole microlens candidate (c.f. the three current
black-hole microlens candidates, which have $\rEt = 29.3,\,11.2~{\rm
 and}~8.7~{\rm au}$; see Agol et al. 2002).
Using equation (\ref{eq:rE}), the mass of the lens for this event is
given by,
\beq \label{eq:mass}
M = 7.21 M_\odot \left( \frac{\tilde{r}_{\rm E}}{\rm10.84\, au}
\right)^2 \left( \frac{\pirel}{0.5 {\rm mas}} \right),
~~\pirel\equiv{{{\rm AU} \over \Dl} - {{\rm AU} \over \Ds} }.
\eeq
A value of $\pirel=0.5$ mas corresponds to a disk source with
$\Ds=2\kpc$ and $\Dl/\Ds=0.5$. There are various approaches that can
be employed to constrain the location of the source and lens, and
hence the lens mass; these are considered in Section \ref{mass}.


To verify the viability of our parallax fit, we proceed to fit the
event with a model that incorporates a constant acceleration term,
instead of the Earth's centripetal acceleration (see Smith, Mao \&
Paczy\'nski 2003). This model is unable to provide a feasible fit for
\sc5, meaning that the parallactic nature of the deviations appears to
be secure.

\section{Constraints on the lens mass}
\label{mass}

\subsection{Finite-source effects}
\label{fss}

Since the peak magnification is predicted to be greater than 40, one
may suspect that this event could be affected by finite-source effects
(Gould 1994; Witt \& Mao 1994; Nemiroff \& Wickramasinghe 1994). Such
effects become apparent when the lens passes sufficiently close to the
source, resulting in an invalidation of the assumption that the source
is point-like. However, there is one significant drawback for \sc5,
namely that there is no coverage for the peak of the light-curve,
i.e. the point at which the lens and source are in closest alignment
and therefore where the finite-source effects should be most prominent.

To implement the finite-source model requires an additional parameter,
$\rho_\ast$, which denotes the source radius in units of the lens'
angular Einstein radius. Despite the lack of coverage around the peak,
constraints can be placed on $\rho_\ast$\footnote{
Formally, it is possible to place a $2\sigma$ lower-limit on
$\rho_\ast$ ($\rho_\ast>0.024$), although we consider this to be
unphysical since it would result in a highly unlikely value of $x=D_{\rm
L}/D_{\rm S}>0.99$.},
\beq
\label{3sig}
\rho_\ast<0.0418~~\mbox{at the 3$\sigma$ confidence level}.
\eeq

This constraint on $\rho_\ast$ can be used to place a lower-limit on the
mass of the lensing object. To do this, we first require an estimate
of the angular size of the source star. From the colour-magnitude
diagram presented in Fig. \ref{fig:CMD}, it appears that the source
lies on the main-sequence branch.
The extinction and reddening for typical bulge stars in this field can
be estimated from the position of the red clump region on the observed
colour-magnitude diagram (see, for example, Albrow et al. 2000). The
location of the centre of the red clump region for this field is given by 
$(V-I)_{\rm cl, obs} \approx 3.82$ mag. Since the intrinsic dereddened
colour of the red-clump region is $(V-I)_{\rm cl, 0} \approx 1.00$ mag
(Popowski 2000), this implies that the clump is reddened by
$E(V-I)_{\rm cl} \approx 2.82$ mag. The slope of the reddening line
for this field ($A_I/E[V-I]=0.95$ for $l=0.1^\circ$, $b=-1.8^\circ$,
Udalski 2002), yields the extinction for the red clump region, 
$A_{I, {\rm cl}} \approx 2.68$ mag. Therefore a star located in the
red clump region, i.e. in the Galactic bulge, will undergo extinction
and reddening of,
\beq
A_{I, {\rm cl}} \approx 2.68~{\rm mag}, ~~ E(V-I)_{\rm cl} \approx
2.82~{\rm mag}.
\label{eq:red}
\eeq

In Section \ref{data} we stated that the observed magnitude and colour
of \sc5 is $I_{\rm obs} \approx 18.5$ mag and $(V-I)_{\rm obs}
\approx 1.97$ mag. Since the best-fit parallax model presented in
Section \ref{model} predicts that there is no blending 
($\fs=1.000^{+ 0}_{-0.065}$, i.e. all of the observed flux comes from
the source), this implies that the observed magnitude and colour of
the source is $I_{\rm S, obs} \approx 18.5$ mag and $(V-I)_{\rm S,
  obs} \approx 1.97$ mag. Therefore, if the source were located in the
bulge (with $D_{\rm S}\approx8$ kpc) and underwent the same amount of
reddening and extinction as given in equation (\ref{eq:red}), the
absolute magnitude and intrinsic colour would be $M_I \approx 1.3$ mag
and $(V-I)_{\rm S, 0} \approx -0.85$ mag. However, this is clearly
incompatible with spectral types known to be in the bulge. Therefore,
we conclude that the source is unlikely to be a Galactic bulge star,
i.e., \sc5 is more-likely an example of disk-disk lensing. By taking a
simple model for the extinction we can obtain the absolute magnitude
and colour of the source as a function of $D_{\rm S}$.
We model the extinction using an exponential dust sheet of
scale height 130pc (e.g. Drimmel \& Spergel 2001), with the Sun located
20pc above the Galactic plane (Humphreys \& Larsen 1995). This simple
analysis suggests that the source is consistent with a K- or G-type
dwarf at a distance of approximately $1.5-2.5$ kpc. For 
example, for $D_{\rm S} \approx 2.0$ kpc, this gives $M_{I,{\rm S}} \approx
6.0$, $I_{\rm S,0}\approx 17.5$ and $(V-I)_{\rm S,0} \approx
0.86$, with $A_I\approx1.1$ mag.

We can check this conclusion by utilising a different approach. If we
assume that the source is a typical G5 main-sequence star with
absolute magnitude $M_I=4.41$ mag and $(V-I)_{\rm C}=0.69$ mag (from Cox 2000,
converted into the Cousins system using Bessell 1979), then, given the
slope of the reddening line for this field $A_I/E[V-I]=0.95$ (Udalski
2002), we can calculate the predicted source distance. This method
suggests that $D_{\rm S}\approx 3.8$ kpc and $A_I\approx1.2$
mag. Applying the same approach to a fainter K5 main-sequence star
with $M_I=6.09$ mag and $(V-I)_{\rm C} = 1.26$ mag (from Cox 2000)
suggests that $D_{\rm S}\approx2.3$ kpc and $A_I\approx0.7$ mag. 

Clearly, the exact brightness and colour of the source will vary
depending on the assumptions; for the following analysis we shall
proceed with the values calculated above using the simple extinction
model and taking $D_{\rm S}\approx 2.0$ kpc. Spectroscopic
observations of the source would be very useful to determine its
spectral-type and hence constrain the distance to the source.

Our prediction for the intrinsic brightness and colour of the source
can be used to estimate its angular size through an available
empirical relationship. For example, van Belle (1999) provides the
following relationship for B- to G-type main-sequence stars with
$-0.4<(V-K)_0<+1.5$ mag,
\beq
\rm{log}~(2 \theta_\ast) + \frac{V_{\rm S, 0}}{5} = 0.500 \pm 0.023
+ (0.264 \pm 0.012) \times [(V-K)_{\rm S, 0}].
\eeq
The source star's intrinsic colour of $(V-I)_{\rm S, 0}\approx0.86$
can be converted into $(V-K)_{\rm S, 0}\approx1.91$ (by using, for
example, Table II of Bessell \& Brett 1988), which implies the star's
angular radius is,
\beq
\theta_\ast \approx 1.08~\mu\rm{as},
\eeq
If the distance to the source is $\sim 2.0$ kpc, this corresponds
to a physical source radius of $\sim 0.5 R_\odot$, i.e. slightly less
than typical K- or G-type dwarfs (e.g. Table 15.8 of Cox [2000] gives
the physical radius of K- and G-type dwarfs to be between
$0.72R_\odot$ and $1.1R_\odot$).

This value for $\theta_\ast$ can be used to estimate the $3\sigma$
lower limit on the lens mass through the formula,
\beq
\label{eq:theta_E}
M = c^2(4G)^{-1}\rEt\theta_{\rm E},
\eeq
since the lens' angular Einstein radius is given by $\theta_{\rm E} =
\theta_\ast/\rho_\ast$. Using the $\rho_\ast$ constraint provided in
equation (\ref{3sig}) gives,
\beq
\label{m3sig}
M>0.034\,M_\odot~~\mbox{at the 3$\sigma$ confidence level},
\eeq
which corresponds to $x=D_{\rm L}/D_{\rm S}<0.995$.

Therefore we are only able to place a weak lower-limit on the lens
mass from finite-source considerations.
This outcome can be understood when one considers the likely values
of the parameter $\rho_\ast$. From equation (\ref{eq:theta_E}), we can
obtain the following relation for $\rho_\ast$,
\beq
\rho_\ast = 0.003 \left( \frac{\rEt}{10.84\,\rm{au}} \right)
\left( \frac{R_\ast}{R_\odot} \right)
\left( \frac{\Ds}{2.0\,\rm{kpc}} \right)^{-1}
\left( \frac{M}{M_\odot} \right)^{-1},
\eeq
where $R_\ast$ is the physical source radius, which we expect to be
$\sim 1R_\odot$ if the source for \sc5 is a main-sequence star.
Gould (1994) showed that the detection of finite-source signatures is
only possible provided $\rho_\ast > 1 / A_{\rm max}$, where $A_{\rm max}$
is the peak magnification. For \sc5 we have $1 / A_{\rm max} \approx
0.023$, which implies that finite-source signatures are extremely
unlikely to be detected.


\subsection{Additional considerations regarding the lens mass}
\label{lhood}

In addition to the above weak constraint placed on the lens mass
from the finite-source analysis, there is other information that we
can use from the properties of the best-fit parallax model,
i.e. independent of the above finite-source fit. First, we shall
consider the parallax velocity parameter ($\tilde{v}$), and then
incorporate our knowledge of the limits on the blended flux.

For parallax microlensing
events the velocity parameter $\tilde{v}$ (i.e. the transverse
velocity of the lens relative to the source, projected onto the
observer plane), must be sufficiently small so that the Earth's
orbital motion is able to affect the light curve. For \sc5,
this transverse velocity is $\tilde{v}=34.2^{+1.5}_{-1.9}~
\rm{km s}^{-1}$ directed almost parallel to the Galactic plane in the
direction of rotation ($\theta_{\rm G}=-6.1$ degrees; see Section
\ref{model}). The velocity vector $\tilde{\bf v}$ is related to the
transverse velocities of the observer (i.e. the Sun,
${\mathbf{v}}_\odot$), source (${\mathbf{v}}_{\rm S}$) and lens
(${\mathbf{v}}_{\rm  L}$), where ${\mathbf{v}}_\odot$,
${\mathbf{v}}_{\rm S}$ and ${\mathbf{v}}_{\rm L}$ are the
2-dimensional velocities perpendicular to the line-of-sight (i.e. in
the lens plane), through the equation,
\beq
\label{eq:vproj}
\tilde{\mathbf{v}}=
\frac{{\mathbf{v}}_{\rm L} - x {\mathbf{v}}_{\rm S}}{1-x}
 - {\mathbf{v}}_\odot.
\eeq

Previous studies of parallax microlensing events (e.g. Alcock et
al. 1995) have used the parameter $\tilde{\bf v}$ and equation
(\ref{eq:vproj}) to obtain a likelihood function for $x$ (and hence,
from equation [\ref{eq:rE}], a likelihood function for the lens mass),
\beq
\label{eq:lhood}
L(x; \tilde{\bf v}) \propto \sqrt{x(1-x)}\, \rho_{\rm L}(x) \,
|\tilde{\bf v}| \, (1-x)^3 \,
\int f_{\rm v, S}({\bf v}_{\rm S}) \, f_{\rm v, L}([1-x][{\bf
  v}_\odot+\tilde{\bf v}]+x{\bf v}_{\rm S})\,d{\bf v}_{\rm S},
\eeq
where $f_{\rm v, L}$ and $f_{\rm v, S}$ are the lens and source velocity
distributions, $\rho_{\rm L}(x)$ is the density of lenses at a
distance $x$, and all vectors are described in the plane perpendicular
to the line-of-sight to the Galactic centre (i.e. two dimensional).
A more thorough analysis may be obtained by including a mass function
prior (see, for example, Agol et al. 2002). However, we do not attempt
such an analysis here;
for the purposes of this work we follow the approach of Bennett et
al. (2002b) and assert that the likelihood function represents all of
our knowledge about the lens mass and location (i.e. we select a
uniform prior), meaning that the above likelihood function can be
interpreted as the probability distribution for $x$.

\begin{figure}
\plotone{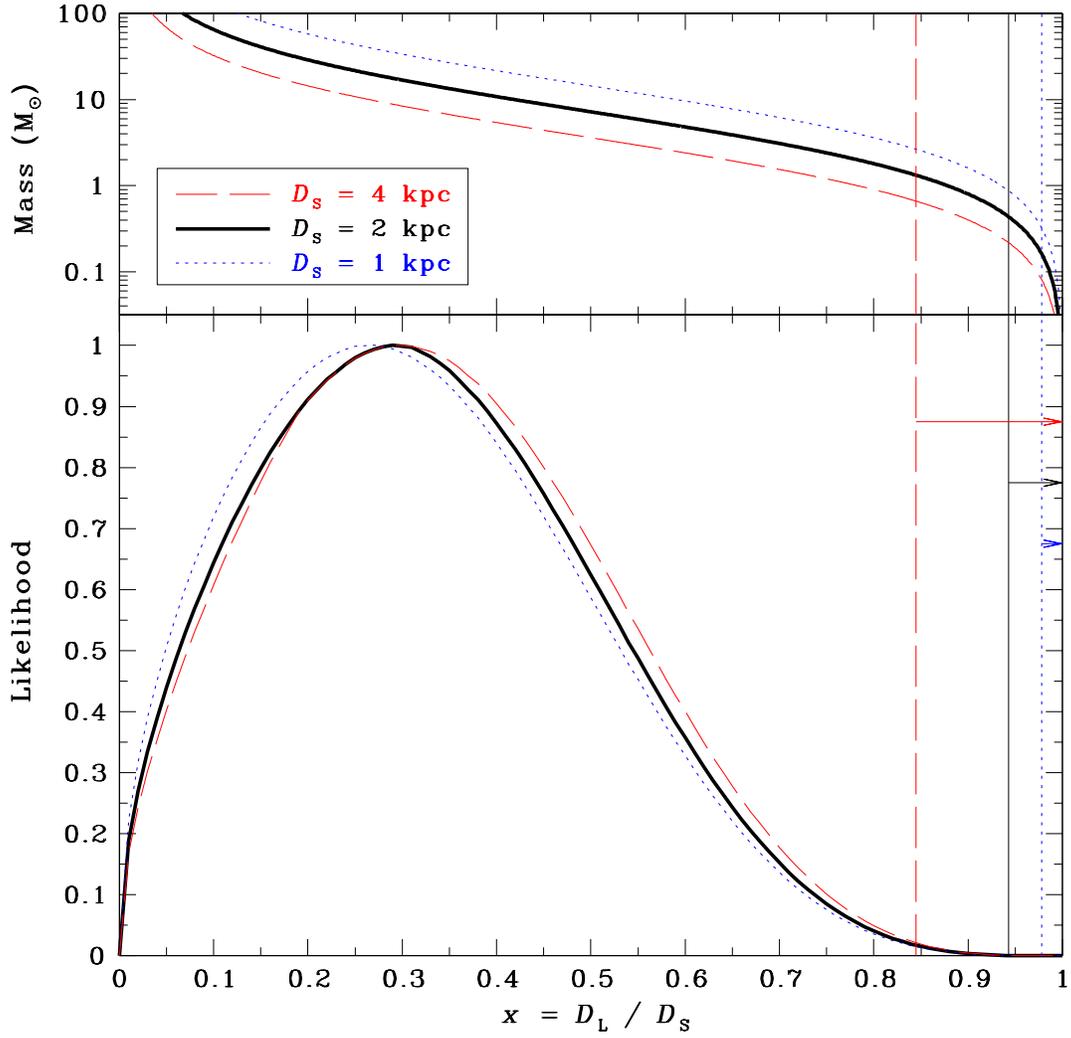}
\caption{
Bottom Panel: The likelihood function for $x=D_{\rm L}/D_{\rm S}$
given the observed parallax velocity parameter ($\tilde{v}=34.2$
km/s in a direction $\theta_{\rm G}=-6.1^\circ$) for event \sc5. As
well as the preferred value of $D_{\rm S}\approx2.0$ kpc (see Section
\ref{fss}), two additional values are presented for comparison. The
vertical lines represent the $2\sigma$ {\it lower} limits on $x$ for
each value of $D_{\rm S}$; these constraints are found by considering
the upper limits on the lens luminosity, assuming that the lens is a
main-sequence star (see Section \ref{lhood}).
Top Panel: This shows how the lens mass varies with $x$ for each value
of $D_{\rm S}$.
}
\label{fig:lhood}
\end{figure}

We evaluate this function by assuming that both the
lens and source reside in the disk, using density and mass
distributions from Belokurov \& Evans (2002) and taking
the Sun's two dimensional peculiar velocity to be 8.89 km/s in a
direction $\theta_{\rm G} = 53.8^\circ$ (Dehnen \& Binney 1998).
The results of this calculation are presented in Figure
\ref{fig:lhood}.
From this figure, we can see that the distribution for
$x$ is not very narrow, implying that the lens can take a wide range
of masses. For example, for $D_{\rm S}\approx 2.0$ kpc, this analysis
suggests that $x=0.32^{+0.20}_{-0.17}$  and hence the lens mass
$M=15.6^{+27.4}_{-8.8} M_{\odot}$; moreover, the $2\sigma$ lower limit
on the lens mass is $3.1 M_\odot$, suggesting that a low mass lens is
strongly disfavoured.

Additional constraints can be placed on the lens nature by
considering the blended flux parameter, $\fs$. The best-fit parallax
model given in Section \ref{model} predicts that there is no blending
(i.e., $\fs=1$), meaning that all of the flux is being emitted by the
lensed source. This is important because it suggests that the lens
may not be a main sequence star, since if this was true one would
expect some blending due to the light from the lens, i.e. one would
expect $\fs<1$.

This possibility can be investigated by considering the limits on the
$\fs$ parameter. From the best-fit parallax model, we can say that the
$2\sigma$ and $3\sigma$ limits on the lens brightness are
$I_{\rm L, obs} > 20.14$ and $I_{\rm L, obs} > 19.62$,
respectively. If we model the lens with a main-sequence
mass-luminosity relation $L \propto M^4$, we can use this to place a
lower limit on $x$ through equation (\ref{eq:rE}). However, to do this
we need to know the source distance, $D_{\rm S}$, and also how the
extinction varies with distance. If we again take $D_{\rm
S}\approx2.0$ kpc and assume the previous simple extinction model (see
Section \ref{fss}), we obtain $2\sigma$ and $3\sigma$ 
limits of $x>0.944$ and $x>0.937$, respectively. This conclusion depends
only weakly on the assumptions regarding the extinction and
mass-luminosity relation.
These constraints on $x$ appear to contradict the above conclusions
from the $\tilde{v}$ likelihood analysis, which suggested that values
of $x$ close to 1 are strongly disfavoured due to the low value of
$\tilde{v}$. Figure \ref{fig:lhood} shows how the $2\sigma$ limits on
the lens brightness compare to the likelihood distribution described
in equation (\ref{eq:lhood}) for three values of $D_{\rm S}$. These
constraints appear to be incompatible with the above likelihood
analysis (for example, for $D_{\rm  S}=2.0$ kpc, the likelihood
analysis gives the probability that $x$ satisfies the $2\sigma$
constraints on the lens brightness to be $<10^{-4}$).

Therefore, we tentatively conclude that the lens for \sc5 is unlikely
to be a main-sequence star, which implies that this event could be
a white dwarf (although this seems unlikely, given that typical white
dwarf masses of $\la 1 M_\odot$ [Bergeron, Ruiz, \& Leggett 1997] are
disfavoured from our likelihood analysis), a neutron star, or possibly
an example of lensing by a stellar mass black hole (Agol et
al. 2002; Bennett et al. 2002a).


\section{Discussion}
\label{disc}

\subsection{The nature of event \sc5}

One approach that can be utilised to probe the nature of the lens for
\sc5 is through proper motion analysis. If the proper motion of the
source can be measured, then this can be combined with the parallax
model's prediction for the transverse velocity of the lens relative to
the source (i.e. $\tilde{v}$) to determine the proper motion of the
lens. The importance of this method is that it can be applied
even in the case where the lens is not luminous. By analysing one
OGLE-II field, Sumi, Eyer \& Wo\'zniak (2003) have shown that proper
motion measurements can be determined for objects in the OGLE-II
catalogue. Preliminary results from the analysis of all OGLE-II
fields suggests that the source star for \sc5 may have a large proper
motion (Sumi, private communication), although this requires
verification as the source star is relatively faint.

Despite the fact that the light curve for \sc5 appears to be
reasonably well fit by the parallax microlensing model (see Section
\ref{model}), it is not inconceivable that the variation in brightness
could be due to some phenomena other than microlensing. Whilst this
ambiguity can affect any microlensing event, for \sc5 it is
particularly severe since data only exist for the declining branch of
the light curve. Indeed, recently Cieslinski et al. (2003) carried
out a search of OGLE-II light curves in an attempt to identify
new cataclysmic variables. In this work they suggested that the shape
of the light curve for \sc5 indicates that this event could
be a candidate nova. However, it is known that microlensing events,
unlike many intrinsically variable stars, are expected to be
achromatic (this may not be true for heavily blended events, but for
\sc5 the best-fit model predicts no blended flux). For \sc5 the
OGLE data only provide colour information for two epochs, namely 
${\rm JD}-2450000\approx 608$ days, and ${\rm JD}-2450000\approx 1020$
days. Reassuringly, the $(V-I)$ colour appears to stay roughly
constant for these two epochs, with $(V-I)=1.97\pm0.02$ and
$(V-I)=2.06\pm0.07$, respectively, which supports our belief that \sc5
is a microlensing event.

\subsection{Scientific returns from long duration events such as \sc5}

An important aspect of \sc5 that is worth noting is that the above
analysis is possible even though data exist for less than half of the
microlensing light curve. Despite the fact that observations
commenced after the peak, it is still possible to identify the
parallactic signatures and gain well-constrained measurements of $\rEt$
and $\theta$. Obviously, this is only possible for \sc5 due to the
high-quality data and the long duration of the event. In future,
even with initially sparse sampling, long-duration events can provide
useful information provided high-quality data are obtained for the
latter part of the light curve. This highlights the importance of
real-time microlensing alert systems (such as the OGLE-III Early
Warning
System\footnote{http://www.astrouw.edu.pl/\~{}ogle/ogle3/ews/ews.html}
or the MOA Transient Alert
Page\footnote{http://www.roe.ac.uk/\~{}iab/alert/alert.html}) and also
the importance of high-quality follow-up observations (such as those
performed by the PLANET collaboration\footnote{http://mplanet.anu.edu.au}).

On the other hand, if real-time alert systems can detect events at
a suitably early stage, then it could be possible to make direct
determinations of the lens mass. For example, once the ESO
Very Large Telescope Interferometer becomes fully operational, its
high sensitivity may enable measurements to be made of the angular
separation of the two microlensed images (Delplancke, G\'orski \& Richichi
2001). Coupled with a measurement of the parallax effect, this would
completely break the lens degeneracy and unambiguously determine the
mass of the lensing object. In future this approach could
prove to be very important for long-duration microlensing events.

Another approach that would benefit from timely follow-up observations
is the method of combining the parallax model with finite-source
effects. As was shown in Section \ref{fss}, if firm measurements can be
made of both parallax and finite-source effects, is it possible to
completely break the lens degeneracy and determine the lens
mass. Although this was not possible for \sc5, it has recently been
shown to be feasible (for example, this approach was applied to the
OGLE-II event sc26\_2218, resulting in a tentative lens mass
determination [Smith, Mao \& Wo\'zniak 2003]).
If real-time alert systems are able to identify high-magnification
long-duration events suitably early, it may be possible to obtain good
quality data for the peak of the light curve and hence determine the
lens mass.

\subsection{An excess of long-duration events in OGLE-II?}
\label{disc:excess}

With the addition of \sc5, this brings the total number of known
events in the OGLE-II catalogue with $\tE>1$ year to three:
OGLE-1999-BUL-32\footnote{
Event OGLE-1999-BUL-32 was independently detected by the MACHO collaboration,
by which it was named MACHO-99-BLG-22. After combining both MACHO and
OGLE datasets, along with additional follow-up data from GMAN (Becker
2000) and MPS (Rhie et al. 1999), they find $\tE = 560 \pm 45$ days
(Bennett et al. 2002b).} ($\tE=640^{+68}_{-54}$ days; Mao et al. 2002),
\sc5 ($\tE=547.6^{+22.6}_{-7.8}$ days; see Section \ref{model} above), and
OGLE-1999-BUL-19 ($\tE=372.0\pm{3.3}$ days; Smith et
al. 2002). Information regarding these three events is presented in
Table \ref{tb:excess}.


\begin{table*}
\begin{center}
\caption{The three longest-duration events in the OGLE-II
catalogue. The fit parameters are derived from the best-fit parallax
model to the OGLE-II Difference Image Analysis data (Wo\'zniak et al.
2001). The parameters $\tE$, $\rEt$ and $\fs$ are defined in Section
\ref{model}, $I_0$ is the total $I$-band baseline magnitude of the
source plus any blended star(s), if present, and $A_{\rm max}$ is the
peak magnification (predicted from the best-fit parallax model). Event
OGLE-1999-BUL-32 was also detected by the MACHO collaboration (by
which it was named MACHO-99-BLG-22) and they find $\tE = 560 \pm 45$
days and $\rEt = 24.3 \pm 3.5$ au (Bennett et al. 2002b).}
\label{tb:excess}
{
\begin{tabular}{ccccccccc}
Name & $\tE$ (day) & $\rEt$ (au) & $I_0$ & $\fs$ & $A_{\rm max}$ & RA (J2000) & Dec
(J2000) & Reference\\
\hline
OGLE-1999-BUL-32 & $640^{+68}_{-54}$ & $29.1^{+6.4}_{-5.4}$& 18.1 & 0.37 & 31.8 & 18:05:05.35 & -28:34:42.5
& Mao et al. (2002)\\
sc5\_2859 & $547.6^{+22.6}_{-7.8}$ & $10.84^{+0.44}_{-0.41}$ & 18.5 & 1.00 & 43.9 & 17:50:36.09 & -30:01:46.6 &
Section \ref{model}\\
OGLE-1999-BUL-19 & $372.0\pm{3.3}$ & $2.68 \pm 0.02$ & 16.1 & 0.82 & 10.2 & 17:51:10.76 & -33:03:44.1 & Smith
et al. (2002)
\end{tabular}
}
\end{center}
\end{table*}

\begin{figure}
\plotone{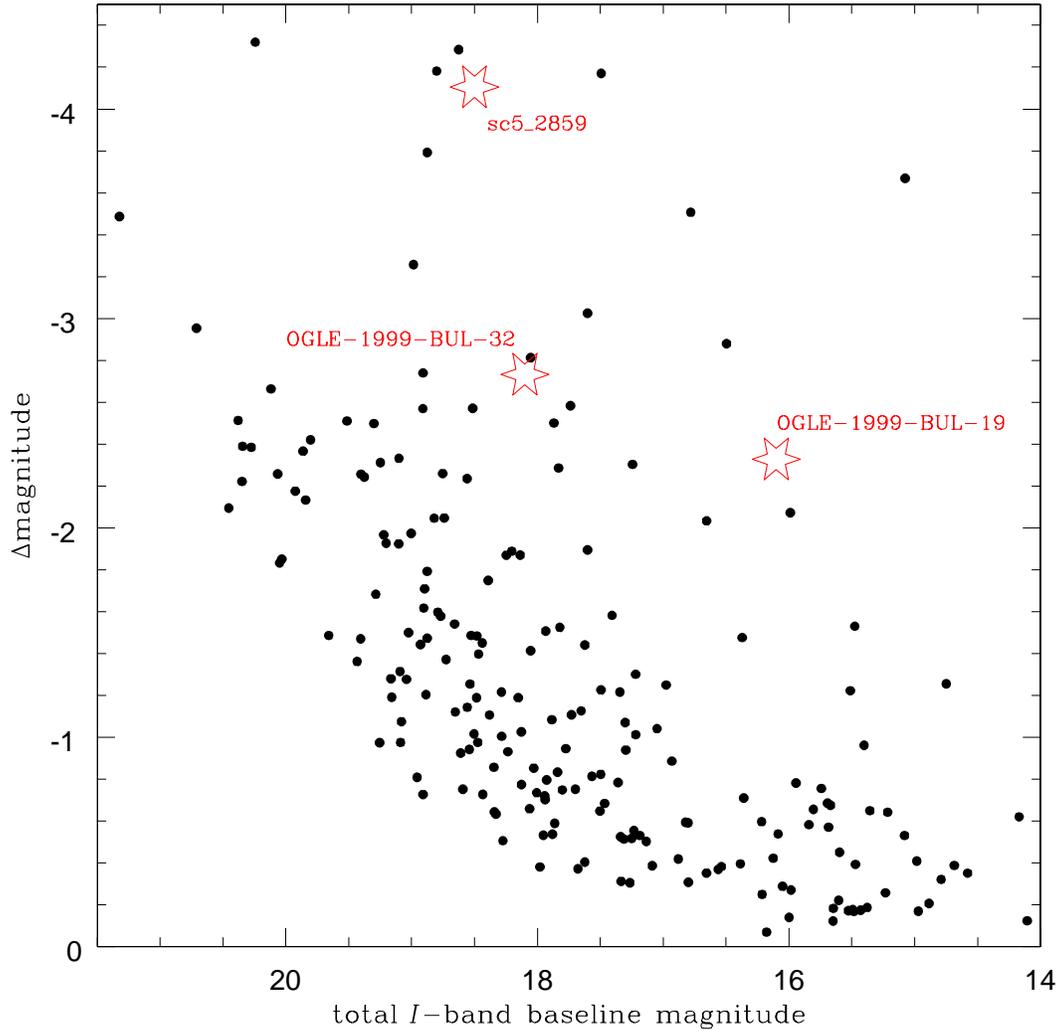}
\caption{The total $I$-band baseline magnitude vs the change in
magnitude at peak magnification (as predicted by the best-fit
standard model) for a high-quality 257-event subset of the OGLE-II
Difference Image Analysis catalogue (Wo\'zniak et al. 2001). The three
longest-duration events are denoted by the star symbols (for these
events the change in magnitude is based on the best-fit parallax
model). Since the change in
magnitude is based on a model fit, it is possible that for some events
(i.e. those for which the peak magnification is not covered by the data)
the predicted change in magnitude may be unreliable. The lack of
events in the lower-left of the diagram is to be expected and is
simply due to the OGLE-II detection limit.
It appears that these three longest-duration events do not lie
within the main cluster of events and this could have important
implications for the abundance of such long-duration events (see
Section \ref{disc:excess}).
}
\label{magcomp}
\end{figure}

Table \ref{tb:excess} shows that these three OGLE-II events are
all highly magnified, with peak magnifications ranging from 10.2 to
greater than 40. Figure \ref{magcomp} illustrates how these
magnifications compare to a high-quality subset of the OGLE-II
catalogue. By plotting the total baseline magnitude vs the change in
magnitude at peak magnification, it can clearly be seen that these
three events do not appear to lie within the main cluster.
This could suggest that some, if not all, of these three long duration
events are artifacts, i.e. the variability may not be due to
microlensing. However, we believe that this is more-likely due to a
selection effect, which implies that there 
could be additional long-duration events residing in the main
lower-magnification cluster, i.e., events that have been omitted
from existing OGLE-II microlensing catalogues due to insufficient
magnification (low signal-to-noise ratio).
In addition, from this table it is interesting to note that that
these events do not appear to be clustered in any one direction
(c.f. Popowski 2002, where it is claimed that an abundance of long
duration MACHO events appear to be clustered in one particular
direction).

%

Previous work has indicated the existence of an excess of long
duration microlensing events toward the Galactic bulge (e.g., Han \&
Gould 1996; Zhao, Rich \& Spergel 1996; Bennett et
al. 2002a). However, is it possible to quantify the extent to which
these three longest-duration OGLE-II events constitute an excess?
If one na\"{\i}vely extrapolates the behaviour of the timescale
distribution of moderately long duration OGLE-II events to long
timescales using a $\tE^{-3}$ power-law (Mao \& Paczy\'nski 1996),
this predicts far fewer than three events should have $\tE>1$
year. However, to perform this calculation thoroughly requires the
detection efficiency of the OGLE-II catalogue, which is currently
undetermined. We aim to investigate this important issue and report
our findings in a future paper.

One implication of this potential excess is that many long-duration
events could easily be overlooked by microlensing search
projects. Microlensing detection algorithms usually rely on the
presence of a constant baseline to differentiate true events from
variable stars. However, such long-duration events may exhibit
noticeable magnification over the course of many years, which can mean
that a constant baseline is not observed. For example, the detection
algorithm of Wo\'zniak et al. (2001) initially failed to identify the
three longest-duration events (i.e., those presented in Table
\ref{tb:excess}) owing to the fact that they did not
pass the constant-baseline criteria. These three events were only
found when this OGLE-II catalogue was compared with a previous,
independently constructed, catalogue of OGLE-II events.
Also, as was mentioned above, it is conceivable
that there are additional long-duration events in the OGLE-II
catalogue that have been omitted due to having insufficient
magnification (i.e., low signal-to-noise ratio).
However, it should be noted that if some long-duration events
are being overlooked in microlensing catalogues then this would
increase the observed optical depth, which is already significantly 
larger than current theoretical predictions (see, for example, Binney,
Bissantz \& Gerhard 2000; Evans \& Belokurov 2002; Klypin, Zhao \&
Somerville 2002).

\section{Conclusion}
\label{conc}

This paper has presented a new long-duration parallax event from
the OGLE-II database, \sc5, which has $\tE = 547.6^{+22.6}_{-7.8}$
days. We believe that both the lens and source
reside in the Galactic disk, making event \sc5 one of the
first confirmed examples of so-called disk-disk lensing. In this
aspect it differs from the two other longest-duration
events, which are both believed to have source stars located
in the Galactic bulge (Mao et al. 2002, Bennett et
al. 2002b; Smith et al. 2002).

The source star for \sc5 is likely to be located at a distance
of $D_{\rm S}\sim 2.0$ kpc, although we are not able to provide a
definitive value (see Section \ref{fss}). This is the major source of
uncertainty in the analysis presented in Section \ref{mass}.
We strongly recommend that spectroscopic observations of
the source are carried out in order to determine the spectral-type of
the source and hence improve the estimate of $D_{\rm S}$.


In Section \ref{lhood} we showed that the transverse velocity
of the lens relative to the source ($\tilde{v}$) can be used to
construct a likelihood function for $x=D_{\rm L}/D_{\rm S}$ and hence
the lens mass.
This analysis suggests that lens is unlikely to be a
main-sequence star, since the resulting luminosity of
the lens would exceed the limits imposed on the blended flux
from the parallax model. Therefore \sc5 could be another
candidate for microlensing by a stellar mass black hole
(Agol et al. 2002; Bennett et al. 2002a).


\section*{Acknowledgements}

This work was supported by a PPARC grant. I would like to thank Shude Mao
for many insightful comments and for detailed criticisms of draft
versions of this manuscript. I am also grateful to Bohdan Paczy\'nski
for helpful suggestions and comments, particularly concerning the
importance of event \sc5 and Fig. \ref{magcomp}, and to Andy Gould for
his advice and guidance.
The full 4-year data for event \sc5 was generously provided by the
OGLE collaboration and I gratefully acknowledge the help of Igor
Soszy\'nski and Przemys{\l}aw Wo\'zniak for their assistance.
In addition, I would like to thank Takahiro Sumi for providing
preliminary data from his proper motion analysis.

\section*{Note added in proof}
After this paper was accepted for publication, it
was discovered that the EROS microlensing collaboration also observed
the field of sc5\_2859 during this event. Preliminary analysis of
these data suggest that they are in good agreement with the OGLE-II data
during the period when the observations overlapped. However, there are
a number of EROS data-points prior to the beginning of the OGLE-II
observations (i.e. for the period $\rm{JD}-245000 < 550$ d), and these
data appear to be incompatible with the microlensing model presented
in this paper. It is hoped that the definitive analysis of the EROS
data, when combined with relevant follow-up observations, will help to
clarify the nature of this event. The results of these findings will
be presented at a later date.

{}

\clearpage

\end{document}